\documentclass[hyper]{JHEP} 

\usepackage{epsfig}

\newcommand\fverb{\setbox\pippobox=\hbox\bgroup\verb}

\newcommand\fverbdo{\egroup\medskip\noindent%

            \fbox{\unhbox\pippobox}\ }

\newcommand\fverbit{\egroup\item[\fbox{\unhbox\pippobox}]}

\newbox\pippobox

\title{Open String in
Non-Relativistic Background}
\author{by J. Kluso\v{n}\\
     Department of Theoretical Physics and Astrophysics\\
                   Faculty of Science, Masaryk University\\
Kotl\'{a}\v{r}sk\'{a} 2, 611 37, Brno\\
Czech Republic\\
    E-mail: \email{klu@physics.muni.cz}}
\preprint{0912.4587}

 \abstract{This note is devoted to the study of the
 open string description of Wilson loops and
quarks in  non-relativistic QFT.}


\def\ba{\mathbf{a}}
\def\bai{\left(\ba^{-1}\right)}

\def\pb #1{\left\{#1\right\}}

\def\mH{\mathcal{H}}

\def\mL{\mathcal{L}}

\newcommand \partt{\partial_\tau}
\newcommand \parts{\partial_\sigma}
\begin{document}
\section{Introduction and Summary}\label{first}
Wilson loops  are  non-local gauge
invariant operators in gauge theory in
which the theory can be formulated.
 Mathematically we define  a Wilson loop as
the trace in an arbitrary
representation $R$ of the gauge group
$G$ of the holonomy matrix associated
with parallel transport along a closed
curve $C$ in space-time. Further, as it
is well known from the times of birth
of AdS/CFT correspondence
\cite{Maldacena:1997re} the Wilson
loops in $N=4$ SYM theory can be
calculated in dual description using
macroscopic strings
\cite{Maldacena:1998im,Rey:1998ik}.
This prescription is based on a picture
of the fundamental string ending on the
boundary of
 AdS along the path $C$ specified by the Wilson loop
operator.

Recently the AdS/CFT correspondence has
been generalized to the description of
some
 non-relativistic strongly
coupled conformal systems \footnote{For
review and extensive list of
references, see
\cite{Hartnoll:2009sz,McGreevy:2009xe}.}.
Non-relativistic conformal symmetry
contains the scaling transformation
\begin{equation}\label{scalrel}
x'^i=\lambda x^i \ , \quad t'=\lambda^z
t \ ,
\end{equation}
where $z$ is a dynamical exponent. In
case of $z=2$ this symmetry is enhanced
to Schor\"{o}dinger symmetry
\cite{Hagen:1972pd,Niederer:1972zz,Henkel:1993sg,
Duval:1990hj,Duval:2009vt,Mehen:1999nd,Son:2005rv,
Leiva:2003kd,Correa:2008bi,Nishida:2007pj}.
It is remarkable that it is possible to
find the gravity dual of these
non-relativistic field theories
\cite{Son:2008ye,Balasubramanian:2008dm}
\footnote{For another solutions that
should be dual to non-relativistic
field theories, see
 \cite{Goldberger:2008vg,Barbon:2008bg,Herzog:2008wg,
Maldacena:2008wh,Adams:2008wt,Kovtun:2008qy,Hartnoll:2008rs,
Schvellinger:2008bf,Mazzucato:2008tr,Rangamani:2008gi,
Alishahiha:2009nm,Martelli:2009uc,Ross:2009ar,Kachru:2008yh,
Blau:2009gd,Yamada:2008if,
Bobev:2009mw,Bobev:2009zf,
Bagchi:2009my,Danielsson:2009gi,Bertoldi:2009vn,
Bertoldi:2009dt,Bertoldi:2009dt,Wen:2008hi,Nakayama:2008qm,Minic:2008xa,
Duval:2008jg,Akhavan:2008ep,Sakaguchi:2009de,Bagchi:2009ke,Alishahiha:2009np,
Bagchi:2009ca,
Volovich:2009yh,Akhavan:2009ns,Donos:2009en,Jeong:2009aa}.}.
The asymptotic metric in this case
reads
\begin{equation}\label{schro}
ds^2=\frac{R^2}{r^2}
(-\frac{dt^2}{r^{2(z-1)}}+ 2dt d\xi +
(dx^i)^2)+\frac{R^2}{r^2} dr^2
 +ds_M^2 \ ,
\end{equation}
where $R$ is characteristic radius of
space-time,  $\xi$ is a
compact light-like coordinate and where
$ ds^2_M$ is the metric of an
appropriate compact manifold which
allows (\ref{schro}) to be a solution
to the supergravity equations of
motion. Since $\xi$ is compact the
associated quantum number is
interpreted as the particle number. As
we said above $z$ that appears in the
scaling relation (\ref{scalrel})is a
critical exponent.
 The usual AdS case
corresponds to $z =1$. On the other
hand non-relativistic Dp-brane
backgrounds are characterized by
dynamical exponents $z=2$
\cite{Maldacena:2008wh,Mazzucato:2008tr}.

The aim of this paper is to study some
properties of non-relativistic $d+1$
dimensional field theories using the
classical string solutions in $d+3$
dimensional dual gravity background
\footnote{For previous works along this
issue, see
\cite{Akhavan:2008ep,Kluson:2009dc}.}.
We consider    
open string that moves with constant
velocity along $\xi$ dimension and that
also extends from the boundary of the
space-time into the bulk, reaches its
turning point and then it goes back to the
boundary. According to the
 bulk/boundary correspondence 
the string action evaluated on this
solution provides potential between
quark anti-quark pair. On the other
hand the fact that the string moves
along $\xi$ direction makes the
analysis more interesting. Explicitly,
 we argue that
open string that it stretched from the
boundary  into the bulk of the
space-time is formal solution of the
string equations of motion since the
action evaluated on it is imaginary for
sufficiently large $r$.
The similar situation occurs in the
context of the thermal CFT/AdS
\cite{Gubser:2006bz,Peeters:2006iu,Herzog:2006gh,Chernicoff:2006hi}.
We find that even in the zero
temperature non-relativistic QFT
 the static object that is described by
stretched string moving with constant
velocity along $\xi$ direction loses
an energy in the process when the
energy and particle number are
transferred from the background
plasma  into the end point of the
string and  then it flows along the
string up infinity. When we consider
$\cap$ shaped string configuration that
moves with constant velocity along
$\xi$-direction we find that there is no
tail along $\xi$ direction 
 with agreement with the 
similar analysis performed in
\cite{Chernicoff:2006hi} in case 
of AdS/CFT correspondence.

We hope that our observation considering
motion of the string along $\xi$ direction
in non-relativistic background demonstrates
that the non-relativistic version of
AdS/CFT correspondence is very rich and
 in some way enigmatic so that it deserves
further study. For example, it would be
certainly very interesting to extend this analysis to
the case of thermal non-relativistic background.

 This paper is
organized as follows. In next section
(\ref{second}) we study the static open
string configuration in
non-relativistic background that
describes Wilson loop in
non-relativistic QFT. In section
(\ref{third}) we study the curved
string in given background that moves
with constant velocity along $\xi$
direction. Finally in appendix we
present the Hamiltonian analysis of
the bosonic string in non-relativistic
background.

\section{Open String Description of
Wilson Line in Non-relativistic Field
Theory}
\label{second}
We begin with the open
string description of the  Wilson loop in
non-relativistic field theory.

  Let us consider  Nambu-Goto
form of the string action
\begin{equation}\label{NGaction}
S=-\frac{1}{2\pi\alpha'} \int d\tau
d\sigma \sqrt{-\det \ba} \ ,
\end{equation}
where
\begin{equation}
\ba_{\alpha\beta}=\partial_\alpha x^M
\partial_\beta x^N g_{MN} \ .
\end{equation}
We  parameterize the string
world-sheet with coordinates
$\sigma^\alpha \ , \alpha,\beta=0,1 \ , \sigma^0=\tau \ ,
\sigma^1=\sigma$. 
Further $x^M=(t,r,\xi,x^1,\dots,
x^d)$
parameterize the embedding of the
string into non-relativistic
 background
\begin{equation}\label{Schr}
ds^2=R^2\left(-\frac{dt^2}{r^{2z}}+2\frac{dt
d\xi}{r^2}
+\frac{dx^idx_i}{r^2}+\frac{dr^2}{r^2}\right)
\ ,
\end{equation}
where $R$ is the characterized
 scale of the background 
and where $\xi$ is a compact light-like
direction.
Finally  $x^i, i=1,\dots,d$ together
with $t$ parameterize the boundary of
the space-time (\ref{Schr}) mapped at
$r=0$.

Our goal is to find 
solutions of the   string 
equation of motion that
provide the dual description of
 Wilson lines in the
boundary non-relativistic quantum field
theory. We also presume
general time dependence of $
\xi$.
We start with
 the equations of
motion for $x^M$ that follow from
(\ref{NGaction})
\begin{eqnarray}
\partial_\alpha [g_{MN}\partial_\beta x^N
\bai^{\beta\alpha}\sqrt{-\det \ba}]-
\frac{1}{2}\partial_M
g_{KL}\partial_\alpha x^K\partial_\beta
x^L \bai^{\beta\alpha}\sqrt{-\det\ba}
=0 \ .
\nonumber \\
\end{eqnarray}
In order to find the static solution it
is convenient to
 fix the world-sheet diffeomorphism 
 by imposing the condition
\begin{equation}\label{sg}
t=x^0=\tau \ , \quad  r=\sigma \ .
\end{equation}
In order to describe Wilson line in
dual field theory with general time
dependence of $
\xi$ we consider
following  ansatz
\begin{equation}\label{xa}
 \xi=\xi(\tau), \quad
x=x(\sigma) \
\end{equation}
so that  (\ref{sg}) together with
(\ref{xa}) imply
\begin{eqnarray}
\ba_{\sigma\sigma}&=&
\frac{R^2}{r^2}+\frac{R^2}{r^2}(\parts
x)^2 \ , \quad  \ba_{\tau\tau}=
-\frac{R^2}{r^{2z}}+2\frac{R^2}{r^2}\partt\xi
 \ .
 \nonumber \\
\end{eqnarray}
Then it is easy to  see that
the  equation of motion for $\xi$
takes the form
\begin{eqnarray}
\partial_\tau[ g_{\xi t}
\bai^{\tau\tau}\sqrt{-\det \ba}]=0
\nonumber \\
\end{eqnarray}
that is solved when
$\partt \xi=v_\xi=\mathrm{const}$. Note also
that for $
\partt \xi=v\_\xi$ the 
equation of motion for  $t$ is
trivially satisfied.
Finally we consider
 the equation of motion for $x$
\begin{eqnarray}
\partial_\alpha[g_{xx}\partial_\beta x
\bai^{\beta\alpha}\sqrt{-\det \ba}]=
\parts \left[g_{xx}\parts
x\sqrt{-\frac{\ba_{\sigma\sigma}}
{\ba_{\tau\tau}}}\right]=0 \   \nonumber \\
\end{eqnarray}
that implies the differential equation
for $x$ in the form
\begin{eqnarray}\label{parx}
\parts x=\pm
\frac{C \sqrt{g_{rr}}}
{\sqrt{g_{xx}^2(-g_{tt}-2g_{t\xi}v_\xi)-C^2
g_{xx}}}=\pm\frac{Cr^{z+1}}{ R^2
\sqrt{1-2v_\xi
r^{2z-2}-\frac{C^2}{R^4} r^{2z+2}}} \ ,  \nonumber \\
\end{eqnarray}
where $C$ is an integration constant.
Let us now analyze this differential
equation.
 Clearly it possesses the
 solution $ x=\kappa, C=0$ that corresponds to the
 string that is extended from the boundary
 at $r=0$ into the bulk space-time and that
 moves with constant velocity $v_\xi$ along
 $\xi$ direction. The configuration
provides the dual description of 
an infinite long Wilson line
extended in time direction. 
 However we will argue
below that even if this ansatz
is the solution of the string
equation of motion  it cannot describe
any physical configuration since the
action evaluated on this ansatz is imaginary.
This is a consequence of the non-zero
motion of the string along $\xi$ direction.
We present the resolution of this puzzle
in the next section.

The second solution of the equation
(\ref{parx}) corresponds to $\cap $ shaped string
that extends from the point $x=-L/2 $ at the boundary $r=0$
into the bulk of the Schr\"{o}dinger
space time until it reaches its turning
point  $r_{max}$ at $x=0$ and then it 
approaches the boundary at $x=L/2$.
 The turning point 
$r_{max}$ is the point when
 $\parts x=\infty $  and from 
(\ref{parx}) we see that
it occurs when the denominator 
vanishes
\begin{equation}\label{rmaxeq}
1-2v_\xi r_{max}^{2z-2}-\frac{C^2}{R^4}
r_{max}^{2z+2}=0 \ .
\end{equation}
The distance $L$ between two
end points of the string is
determined 
 by the integral
\begin{eqnarray}\label{Lint}
L=2\int_0^{r_{max}}d\sigma \parts x=
\frac{2 C}{R^2}
\int_{0}^{r_{max}} dr \frac{r^{z+1}} {
\sqrt{1-2v_\xi r^{2z-2}-\frac{C^2}{R^4}
r^{2z+2}}} \ .  \nonumber \\
\end{eqnarray}
In principle
this formula allows  to express
 $r_{max}$ as a function of $L$.
Then we can evaluate the action on
given solution  and find the potential
 $V(L)$
between quark and  anti-quark-like objects
that correspond to  the
 end points of the string. However   in
order to find finite potential $V(L)$ we
have to  subtract the
contribution from two infinite extended
strings, following the standard
recipe
\cite{Maldacena:1998im,Rey:1998ik}.
Implementing this procedure we 
find
\begin{eqnarray}\label{Ssol1}
S&=&-\int d\tau V(L)=
-\frac{2}{2\pi\alpha'}
\int d\tau [\int_0^{r_{max}}
d\sigma \sqrt{-\det \ba}
+\int_0^\infty
\sqrt{-\det \ba}_{st}]
=\nonumber \\
&=&-\frac{R^2}{\pi\alpha'}
\int d\tau \left[ \int_0^{r_{max}} d\sigma
\frac{\sqrt{1-2r^{2(z-1)}v_\xi}}
{r^{z+1}}\left(\frac{1}{\sqrt{1-2v_\xi r^{2z-2}
-\frac{C^2}{R^4}r^{2z+2}}}-1\right)-\right.
\nonumber \\
&  &\left.-\int_{r_{max}}^{\infty}dr
\frac{\sqrt{1-2r^{2(z-1)}v_\xi}}{r^{z+1}}\right] \ ,
\nonumber \\
\end{eqnarray}
where $\sqrt{-\det \ba}_{st}=\frac{1}{r^{(z+1)}}
\sqrt{1-2r^{2(z-1)}v_\xi}$
is determinant evaluated on
the solution of the equation of motion corresponding
to the straight string $x=const$.
However from (\ref{Ssol1}) we
immediately see that the action becomes
imaginary for
\begin{equation}
r^{2(z-1)}>r_*^{2(z-1)}=\frac{1}{2v_\xi}
\ 
\end{equation}
and hence solution corresponding to straight
string is unphysical. Note that
the similar behaviour
was observed in case of
AdS/CFT correspondence
\cite{Gubser:2006bz,Peeters:2006iu,Herzog:2006gh,Chernicoff:2006hi}.
On the other hand using $r_*$ we can rewrite the
 equation (\ref{rmaxeq}) in the form
 \begin{equation}
 1-\left(\frac{r_{max}}{r_*}\right)^{2(z-1)}-
 \frac{C^2}{R^4}r^{2z+2}_{max}=0
 \end{equation}
that implies that  for $r_{max}<r_*$
and hence  $\cap$ shaped string
is well defined since the action
evaluated on this solution is real. 
However as we argued above  in order to regularize 
the divergences coming from the region
around $r=0$  we have to find physical
open string configuration that describes
single quark. We return to this problem
in next section. 

Let us now consider the case when
$v_\xi=0$ so that the equation
(\ref{rmaxeq}) has the solution
\begin{equation}
r_{max}=\frac{R^{\frac{2}{z+1}}}
{C^{\frac{1}{z+1}}} \ .
\end{equation}
It is convenient to introduce the
variable $y=\frac{1}{r}$ so that
$y_{min}^{z+1}=\frac{C}{R^2}$. Then
(\ref{Lint}) can be explicitly
evaluated
\begin{eqnarray}\label{Lym}
L&=&2\int_{y_{min}}^\infty
\frac{dy}{y^2}\frac{y^{z+1}_{min}}{
\sqrt{y^{2(z+1)}-y^{2(z+1)}_{min}}}=
\nonumber \\
&=&\frac{2}{y_{min}}\int_1^\infty dk
\frac{1}{k^2\sqrt{k^{2(z+1)}-1}}=
\frac{2}{y_{min}}
\frac{\sqrt{\pi}\Gamma(\frac{z+2}{2z+2})}
{\Gamma(\frac{1}{2+2z})} \ .
\nonumber \\
\end{eqnarray}
This result allows us to express
$y_{min}$ as a function of $L$. 
Further from (\ref{Ssol})
we find the quark anti-quark potential
$V(L)$
 \begin{eqnarray}
S&=&-\int d\tau V(L)=\nonumber \\
&=&
-\frac{R^2}{\pi\alpha'} \int d\tau
\left[ \int_0^{r_{max}} dr
\frac{1}
{r^{z+1}}\left(\frac{1}{\sqrt{1-(\frac{r}{r_{max}})^{2(z+1)}}}
-1\right) -\frac{1}{r_{max}^zz }\right]=
\nonumber \\
&=&\int d\tau
\frac{R^2}{\sqrt{\pi}\alpha'z}
\left(\frac{2\sqrt{\pi}}{L}\right)^z
\left(\frac{\Gamma(\frac{z+2}{2z+2})}
{\Gamma(\frac{1}{2+2z})}\right)^{z+1} \
,
\nonumber \\
\end{eqnarray}
where in the final step
of calculation we used (\ref{Lym}).
In fact, this result reproduces the 
calculation of the Wilson loop
in the Lifshitz space-time that
was performed in
\cite{Danielsson:2009gi}.

\section{Curved Moving String}\label{third}
In the previous section we argued that 
straight string moving along $\xi$ direction
with velocity $v_\xi>0$ is not well
defined since in
this case the string action becomes
imaginary for  $r>r_*=\left(\frac{1}{2v_\xi}
\right)^{\frac{1}{2(z-1)}}$. The similar
situation occurs in case of the open 
 string in
thermal $AdS_5$ background
\cite{Gubser:2006bz,Peeters:2006iu,Herzog:2006gh,Chernicoff:2006hi}
\footnote{Similar analysis was performed in
\cite{Fadafan:2009an} for  string  in
Lifshitz space-time.}.
 With
analogy with this case
we show
that it is possible to find the
solution of the string equation of
motion that is real and that describes
an infinite long open  string that is stretched
from the boundary at $r=0$ into the bulk of
the space-time. To do this we
propose following ansatz
\begin{equation}
r=\sigma \ , \quad  t=\tau \ , \quad
x=\mathrm{const} \ , \quad  \xi=\xi=v_\xi\tau+y(\sigma)
 \
\end{equation}
so that
\begin{eqnarray}
\ba_{\tau\tau}=g_{\tau\tau}+2g_{t\xi}v_\xi \ ,
\quad
\ba_{\tau\sigma}=\ba_{\sigma\tau}=
g_{t\xi}\parts y
\ , \quad
\ba_{\sigma\sigma}=g_{rr} \ .
\nonumber \\
\end{eqnarray}
Let us now solve the equation of motion
for this ansatz. First of all the
equation of motion for $t$ implies
following differential equation  for $y$
\begin{eqnarray}\label{eqtc}
(\parts y)=\pm
K\frac{\sqrt{-g_{rr}(g_{tt}+2g_{t\xi}v_\xi)}}
{\sqrt{g^4_{t\xi} v_\xi^2-K^2
g_{t\xi}^2}}=\pm
K\frac{\sqrt{\frac{1}{r^{2z-2}}-2 v_\xi}}
{\sqrt{\frac{R^4}{r^4}v_\xi^2-K^2}} \ .
\nonumber \\
\end{eqnarray}
In the same way the equation
of motion for $
\xi$ implies 
\begin{equation}\label{xieqc}
\frac{g_{t\xi}^2 \parts
y}{\sqrt{-\det\ba}}=K_\xi 
 \ , 
 \end{equation}
where $K_\xi$ is a constant. Solving
this equation for $\parts y$ we find
\begin{equation}
\parts y=\pm K_\xi
\frac{\sqrt{-g_{rr}(g_{tt}+2g_{t\xi}v_\xi)}}
{\sqrt{g_{t\xi}^4-K_\xi^2g_{t\xi}^2}} 
  \ .  
\end{equation}
Comparing this equation with 
(\ref{eqtc}) 
 we find the relation between the integration
constant $K_\xi$ and $K$ in the form $K=K_\xi
v_\xi$.

 We see that there is a
possibility that the denominator
becomes imaginary. In order to avoid
this possibility we have to demand that
the denominator and numerator vanish at
the same point. This happens at
\begin{eqnarray}\label{Kcoos}
\frac{1}{r^{2z-2}_*}= 2v_\xi \ ,
\quad
K^2=R^4 2^{\frac{2}{z-1}} v_\xi^{\frac{2}{z-1}} \ .
\nonumber \\
\end{eqnarray}
Then using (\ref{Kcoos}) the equation
(\ref{eqtc}) takes the form
\begin{eqnarray}
\parts y=\pm
\frac{K r^2}{R^2 v_\xi r^{z-1}}
\frac{\sqrt{1-2 v_\xi r^{2z-2}}} {\sqrt{1-
2^{\frac{2}{z-1}}v_\xi^{\frac{2}{z-1}}
r^4}}
 =\pm 2^{\frac{1}{z-1}}v_\xi^{\frac{1}{z-1}}
r^{3-z} \frac{\sqrt{1-2 v_\xi
r^{2z-2}}} {\sqrt{1-
2^{\frac{2}{z-1}}v_\xi^{\frac{2}{z-1}}
r^4}} \ .
\nonumber \\
\end{eqnarray}
Observe that for the most interesting
case $z=2$ this equation simplifies as
\begin{equation}
\parts y
 =\pm 2v_\xi
r \frac{\sqrt{1-2 v_\xi r^{2}}}
{\sqrt{1- 2^{2}v_\xi^{2} r^4}}=
\pm\frac{2 v_\xi r}
{\sqrt{1+2v_\xi r^2}}
 \ 
\nonumber \\
\end{equation}
so that it can be
 be easily integrated
with the result
\begin{equation}
y(r)=\pm\sqrt{1+2v_\xi r^2}\mp 1 \ ,
\end{equation}
where the value of the integration
constant was determined by
the requirement that  for $r=0$ the time
dependence of the end point of the string
is $\xi=\tau v_\xi$.

Let us now return to general
$z$ and
define following currents
\begin{eqnarray}\label{picur}
\pi_t^\alpha&=&\frac{\delta S}{\delta\partial_\alpha t}
=-\frac{1}{2\pi\alpha'}g_{tN}\partial_\beta x^N
\bai^{\beta\alpha}\sqrt{-\det\ba} \ , \nonumber \\
\pi_i^\alpha&=&\frac{\delta S}{\delta\partial_\alpha x^i}
=-\frac{1}{2\pi\alpha'}g_{iN}\partial_\beta x^N
\bai^{\beta\alpha}\sqrt{-\det\ba} \ , \nonumber \\
\pi_\xi^\alpha&=&\frac{\delta S}{\delta\partial_\alpha \xi}
=-\frac{1}{2\pi\alpha'}g_{\xi N}\partial_\beta x^N
\bai^{\beta\alpha}\sqrt{-\det\ba} \ , \nonumber \\
\end{eqnarray}
where $N=(t,x^i,\xi)$. Since the action
does not explicitly depend on $t,\xi$ and $x^i$
it turns out that the currents (\ref{picur})
are conserved
\begin{eqnarray}\label{corcur}
\partial_\alpha  \pi_\xi^\alpha=0 \ ,
\quad \partial_\alpha \pi_t^\alpha=0 \ ,
\quad \partial_\alpha \pi_i^\alpha=0
\ .
\nonumber \\
\end{eqnarray}
In our case the current
$\pi_\xi$ takes the form
\begin{eqnarray}
\pi_\xi^\tau&=&
\frac{1}{2\pi\alpha'}
\frac{g_{t\xi}g_{rr}}{\sqrt{-\det\ba}}=
\frac{1}{2\pi\alpha'}\frac{K_\xi}{
\parts \xi} \ ,
\nonumber \\
\pi_\xi^\sigma&=&
-\frac{1}{2\pi\alpha'}\frac{g_{\xi
t}^2 v_\xi\parts y} {\sqrt{-\det\ba}}=-
\frac{K}{2\pi\alpha'} \ .
 \nonumber \\
\end{eqnarray}
For further purposes we also determine
the spatial component of the current $\pi_t^\sigma$
\begin{eqnarray}\label{sigmat}
\pi_t^\sigma
=\frac{1}{2\pi\alpha'}
\frac{g^2_{t\xi}\parts y
v_\xi}{\sqrt{-\det\ba}} =\frac{K_\xi
v_\xi}{2\pi\alpha'} \ .
\nonumber \\
\end{eqnarray}
Note that in all these calculations
we used
\begin{equation}\label{detba}
\sqrt{-\det\ba}=\frac{g_{t\xi}^2  \parts y}{K_\xi}
\
\end{equation}
that follows from (\ref{xieqc}).

Now we give the physical
interpretation of given solution. Since
 the open string   obeys
the Neumann boundary conditions we see
that there should exist the transport
of the energy and momenta from one end
point of the string to the second one.
Explicitly, let us define the total
energy $E$ and momentum $P_\xi$  as
\footnote{Note that from the point of 
view of dual non-relativistic QFT $P_\xi$ is
interpreted as the operator of number of
particles rather then operator of momenta $P_\xi$.}
\begin{equation}
E=-\int_0^\infty d\sigma \pi^\tau_t \ ,
\quad P_\xi=\int_0^\infty d\sigma
\pi^\tau_\xi \ \ .
\end{equation}
Then  (\ref{corcur}) imply
\begin{eqnarray}
\frac{dE}{dt}&=& -\int_0^\infty d\sigma
\partial_\tau \pi^\tau_t=
\int_0^\infty d\sigma \partial_\sigma
\pi^\sigma_t= \nonumber \\
&=&\pi^\sigma_t(\infty)-\pi^\sigma_t(0)
\ .
\nonumber \\
\end{eqnarray}
Using (\ref{sigmat}) we see that
$\pi^\sigma_t(\infty)=\pi^t_\sigma(0)$
and consequently  the total change of
the energy of the string is zero. On
the other hand we can study the local
gain and lose of the energy at the end
point of the interval. In other words
the string will either lost or gain  an
energy at $r=0$ equal to
\begin{equation}
\left.\frac{dE}{dt}\right|_{r=0}
=-\pi^\sigma_t(0) =
-\frac{K}{2\pi\alpha'}= \mp
\frac{1}{2\pi\alpha'} R^2
2^{\frac{1}{z-1}}v_\xi^{\frac{1}{z-1}}
\ ,
 \
\end{equation}
and at the same time it will either
gain or lost  the energy at $r=\infty$.
Further, the change of  the momentum $P_\xi$
is equal to
\begin{eqnarray}
\frac{dP_\xi}{dt }&=& \int_0^\infty
d\sigma
\partt \pi^\tau_\xi= -\int_0^\infty
d\sigma
\parts\pi^\sigma_\xi=
\nonumber \\
&=&-\pi^\sigma_\xi(\infty)+
\pi^\sigma_\xi(0) \nonumber \\
\end{eqnarray}
and  it will either  gain or lose
 momentum $P_\xi$ at $r=0$  given by
the equation
\begin{equation}
\left.\frac{d P_\xi}{dt}\right|_{r=0}=
\pi^\sigma_\xi(0)=- \frac{K
}{2\pi\alpha'} = \mp
\frac{1}{2\pi\alpha'} R^2
2^{\frac{1}{z-1}}v_\xi^{\frac{1}{z-1}}
\  .
\end{equation}
The concrete situation whether string
gains or lose an energy is determined
 by the sign in front of the constant
 $K$. Note that in the  case of the
 calculation of drag
 force in the context
of $AdS/CFT$
correspondence authors gave  strong
physical arguments for the $-$
 sign in front of $K$\footnote{
More precisely, if $K$ were positive
then the  energy would  flow down
the string toward the horizon while  if
$K$ were negative then the energy would
 flow upward from the
horizon and the tail of string leads
the quark. However it was argued in
\cite{Gubser:2006bz, Herzog:2006gh}
that the physical realistic solution is
the one corresponding to the energy
flow from the quark to the black hole
horizon.}. However since we consider
the background that is dual to
quantum field theory at zero temperature
there is no horizon and it is not completely
clear which sign to choose.
Further, the motion of string
along $\xi$ dimension has different physical
interpretation in the dual QFT
then the motion along one of the
boundary dimensions labelled by $x^i$.
In fact,  strong arguments
were given in
\cite{Adams:2008wt} for the interpretation
of the total momentum along $\xi$ dimension
as the quantum number that counts the
number of particles in given theory.
Then
we suggest to interpret the previous
result as the  energy  and particle flow
along the string  from  its
end point localized at the boundary
$r=0$   towards to the
infinity. Note also   that the
energy and particle flow coincide.
As we said above the boundary
conditions also allow  an existence of the second
kind of  string that describe
the opposite situation when the string
loses an energy and particle number
that are injected into non-relativistic
plasma.
Finally we evaluate the Lagrangian
density for given configuration. Using
(\ref{detba}) we find
\begin{eqnarray}\label{Ldans}
\mL=-\frac{1}{2\pi\alpha'}
\frac{g_{t\xi}^2\parts y}{K_\xi}=
\mp\frac{1}{2\pi\alpha'}
\frac{g_{t\xi}\sqrt{-g_{rr}(g_{tt}+
2g_{t\xi}v_\xi})} {\sqrt{g_{t\xi}^2
-K^2_\xi}} \ .
\nonumber \\
\end{eqnarray}
We see that for $r\rightarrow 0$ the
Lagrangian density (\ref{Ldans})
approaches to
\begin{equation}\label{mLcurv}
\mL=\mp\frac{1}{2\pi\alpha'}
\frac{R^2}{r^{z+1}}
\end{equation}
and hence the action of the string
diverges as $\mp\frac{1}{2\pi\alpha'}
\frac{R^2 }{z \epsilon^z}$ in the limit
of $\epsilon\rightarrow 0$, where the origin
of this divergence is the same as in
AdS/CFT correspondence
\cite{Maldacena:1998im,Rey:1998ik}. Now due 
to the fact that (\ref{mLcurv}) is real
for all $r$ we see that this Lagrangian
density can be imposed as the regularizator
of $\cap$ shaped string. In other words
the quark anti-quark potential 
is determined by following prescription
\begin{eqnarray}\label{Ssol}
 V(L)=\frac{2}{2\pi\alpha'}
\left[\int_0^{r_{max}}
d\sigma \sqrt{-\det \ba}
-\int_0^\infty
\sqrt{-\det \ba}_{tailed}\right] \ .
\nonumber \\
\end{eqnarray}

\section{Wilson Loop Along $x$ Direction
and Tailed String}
In this section return to the situation
studied in the first section when we consider
the  open string description of the
Wilson loop along $x$
direction. We also consider
the situation when the 
string moves with the 
constant velocity $v_\xi$ along
$\xi$  direction. Explicitly
we consider
  following  ansatz
\begin{equation}\label{qantiqa}
\tau=t \ , \quad r=\sigma \ , \quad
x=x(\sigma) \ , \quad
\xi=v_\xi\tau+y(\sigma) \
\end{equation}
that leads to the following
 matrix components
$\ba_{\alpha\beta}$
\begin{eqnarray}
\ba_{\tau\tau}=g_{tt}+2g_{t\xi}v_\xi \ , \quad
\ba_{\tau\sigma}=\ba_{\sigma\tau}=g_{t\xi}\parts
y  \ , \quad
\ba_{\sigma\sigma}=g_{rr}+g_{xx}(\parts
x)^2 \ .
\nonumber \\
\end{eqnarray}
Let us now solve the equations of
motion for $t,\xi$ and $x$. The
equation of motion for $t$ and $\xi$
take the same form as in previous
section and imply
\begin{eqnarray}\label{eqxxi}
\frac{g^2_{t\xi}v_\xi \parts
y}{\sqrt{-\det\ba}} =K \ , \quad
\frac{g^2_{\xi t}\parts
y}{\sqrt{-\det\ba}}= K_\xi \ .
\nonumber \\
\end{eqnarray}
Finally the equation of motion for $x$
implies
\begin{eqnarray}\label{eqx2}
\frac{g_{xx}\parts x \ba_{\tau\tau}}
{\sqrt{-\det\ba}}=K_x \ , \nonumber \\
\end{eqnarray}
where $K_x$ is an integration constant.

We again see that the
equation of motion for $t$ and $\xi$
are proportional so that it is
sufficient to consider one of them, say
the equation of motion for $\xi$. To
proceed further we rewrite
(\ref{eqxxi}) and (\ref{eqx2}) into the
form
\begin{eqnarray}
& &(\parts y)^2 g_{t\xi}^2(g_{t\xi}^2
-K_\xi^2) +K_\xi^2 g_{xx}(\parts x)^2+
K_\xi^2\ba_{\tau\tau} g_{rr}=0 \ ,
\nonumber \\
& &(g_{xx}^2 \ba_{\tau\tau}^2+K_x^2
\ba_{\tau\tau}g_{xx})(\parts x)^2-
K_x^2 g_{t\xi}^2 (\parts y)^2+ K_x^2
\ba_{\tau\tau}g_{rr}=0 \ .
\nonumber \\
\end{eqnarray}
Solving this system of equations
with respect to $\parts x, \parts y$ 
we find
\begin{eqnarray}\label{partsy}
(\parts y)^2&=&-
\frac{K_\xi^2\ba_{\tau\tau}g_{rr}[\ba^2_{\tau\tau}
g_{xx}+\ba_{\tau\tau}K_x^2-K_x^2]}
{[\ba_{\tau\tau}^2g_{xx}+K_x^2\ba_{\tau\tau}]
(g_{t\xi}^4 v_\xi^2-g_{t\xi}^2 K_\xi^2
)+K_\xi^2 K_x^2 g_{t\xi}^2} 
\  , \nonumber \\
(\parts x)^2&=&-\frac{1}{g_{xx}}
\frac{\ba_{\tau\tau}g_{rr}g_{t\xi}^4v_\xi^2
K_x^2} {
[\ba_{\tau\tau}^2g_{xx}+K_x^2\ba_{\tau\tau}]
(g_{t\xi}^4 v_\xi^2-g_{t\xi}^2 K_\xi^2
)+K_\xi^2 K_x^2  g_{t\xi}^2} \ .
\nonumber \\
\end{eqnarray}
We see that $\parts y$ vanishes for
 $\ba_{\tau\tau}=0$ at
\begin{equation}
\frac{1}{r^{2z-2}_*}=2v_\xi \ .
\end{equation}
On the other the turning point
of the Wilson loop defined
by condition  $\parts x=\infty$
occurs when
\begin{equation}
[\ba_{\tau\tau}^2(r_{min})
g_{xx}+K_x^2\ba_{\tau\tau}(r_{min})](g_{t\xi}^2
v_\xi^2- K_\xi^2 )=K_x^2 \ .
\end{equation}
 Further, the consistency
of the solution demands that
 $r_{min}<r_*$. Finally we have to
demand that  the
projections of the two halves of the string onto the
$ \xi - x$ plane have
 to join smoothly so that we have to
require that $\frac{\partial
x}{\partial y}= \frac{\parts x}{\parts
y}=\infty$. Then using (\ref{partsy}) 
 we obtain
\begin{eqnarray}
\frac{\parts x}{\parts y}
&=&\frac{g_{t\xi}^2v_\xi
K_x}{K_\xi}\frac{1}{\sqrt{g_{xx}
(\ba^2_{\tau\tau}
g_{xx}+\ba_{\tau\tau}K_x^2-K_x^2)}} \ .
\nonumber \\
\end{eqnarray}
We see that this expression diverges
for $K_\xi=0$ which however implies 
$\parts y=0$. In other words there is
no tail in case of $\cap$ shaped string
that moves  around  $\xi$ direction with
the velocity $v_\xi$. This
result is in agreement with the similar
analysis performed in case of $\cap$ shaped
 string in AdS/CFT background
\cite{Chernicoff:2006hi}.



\vskip .2in \noindent {\bf
Acknowledgements:} This work was
 supported by the Czech
Ministry of Education under Contract
No. MSM 0021622409. 
I would like also
thank to Max Planck Institute at Golm
for its kind hospitality during my work on
this project.

\begin{appendix}
\section{Wilson loops in Hamiltonian
Formalism} In this appendix we develop
 the Hamiltonian formalism for the open 
string in non-relativistic background
and reproduce the equations derived in
the second section \footnote{The general
 analysis of Hamiltonian
 formulation  of bosonic string in
 non-relativistic background 
 was given in
\cite{Kluson:2009dc}.}.

To begin with we consider the
Nambu-Goto action for 
 string in general background
 \begin{equation}
 S=-\frac{1}{2\pi\alpha'}
 \int d\tau d\sigma 
 \sqrt{-\det \ba} \ 
 \end{equation}
 so that the momenta conjugate
 to $x^M$ are equal to
\begin{equation}\label{pmdef}
p_M=\frac{\delta S}{\delta \partt x^M}
=
\frac{1}{2\pi\alpha'}
\frac{g_{MN}}{\sqrt{-\det\ba}}
(\partial_\tau x^N \ba_{\sigma\sigma}-
\partial_\sigma x^N \ba_{\tau\sigma}) \ .
\end{equation}
Using this expression it is easy to 
derive following first class
constraints
\begin{equation}\label{mHT}
\mH_T\equiv p_M g^{MN}p_N+
\frac{1}{(2\pi\alpha')^2}\parts x^N
g_{MN}\parts x^N\approx 0 \ 
\end{equation}
and 
\begin{eqnarray}
\mH_S\equiv p_M\parts x^M\approx 0 \ .
\nonumber \\
\end{eqnarray}
Since the bare  Hamiltonian density
$\mH_0=\partt X^Mp_M-\mL$ vanishes for
the Nambu-Goto string we find that the
the total Hamiltonian density is
sum of the constraints
\begin{equation}
\mH_T=N \mH_T+N_S \mH_S \ ,
\end{equation}
where $N,N_S$ are Lagrange multipliers.
It is also important to stress that
$\mH_T,\mH_S$ are the first class
constraints and that the consistency of
the time evolution of these constraints
do not generate the secondary ones.

In order to study the static
configuration of string it is
convenient  to fix the gauge symmetries
generated by $\mH_T,\mH_S$. This can be
done  with the help of the gauge fixing
functions
\begin{equation}\label{gff}
G_H: x^0-\tau=0 \ , \quad G_S:
r-\sigma=0 \
\end{equation}
that have non-zero Poisson brackets
with $\mH_T,\mH_S$ and consequently
the set of constraints
$\mH_T,\mH_S,G_T,G_S$ form the
collection of the second class
constraints. It is well known that
these constrains strongly vanish and
can be explicitly solved. In fact,
solving the constraint $\mH_S=0$ for
$p_r$ we find
\begin{equation}\label{pr}
p_r=-p_i\parts x^i-p_\xi \parts \xi \
\end{equation}
using the fact that $\parts r=1,\parts
t=0$ as follows from (\ref{gff}). Then,
inserting this result into $\mH_T=0$
and using (\ref{gff}) we can generally
express $p_0$ as function of remaining
canonical variable.

Let us now specialize to the case of
string in non-relativistic  background.
The characteristic property of this
background is the presence of non-zero
metric component  $g_{t\xi}\neq 0$.
More precisely, let us consider the
background (\ref{Schr}) and determine
corresponding inverse metric
\begin{equation}
g^{tt}=0 \ , \quad
g^{t\xi}=\frac{1}{g_{t\xi}} \ , \quad
g^{\xi\xi}=-\frac{g_{tt}}{g_{t\xi}^2} \
, \quad g^{rr}=\frac{1}{g_{rr}}  ,
\quad
 g_{ij}=\frac{1}{g_{ii}}\delta_{ij} \ .
\end{equation}
Using this form of the metric it is
easy to solve the constraint $\mH_T=0$
for $p_0$ and we find
\begin{eqnarray}
& &\mH_{fix}\equiv -p_0
 =\frac{1}{2g^{t\xi}p_\xi}
 \left(p_\xi g^{\xi\xi}p_\xi+
 (p_i\parts x^i+p_\xi
 \parts \xi) g^{rr}
(p_j\parts x^j+p_\xi\parts
\xi)+\right.\nonumber \\
&+& \left. p_i g^{ij}p_j+
 \frac{1}{(2\pi\alpha')^2}
 (g_{rr}+\parts x^ig_{ij}\parts x^j)\right) \
 ,
\nonumber \\
 \end{eqnarray}
where we used (\ref{Schr}) and
performed the standard identification
between the Hamiltonian density
$\mH_{fix}$ of the gauge fixed theory
and the momentum $p_0$.

Using $H_{fix}$ it is easy to determine
the equations of
 motion for  $\xi$ and
$x^i$
\begin{eqnarray}\label{eqxi}
\partt \xi&=&\pb{\xi,H_{fix}}=\nonumber
\\
&-&\frac{1}{2g^{t\xi}p_\xi^2}
\left(p_\xi g^{\xi\xi}p_\xi+
 (p_i\parts x^i+p_\xi
 \parts \xi) g^{rr}
(p_j\parts x^j+p_\xi\parts \xi)
 +p_i g^{ij}p_j+\right.\nonumber \\
 &+&
\left. \frac{1}{(2\pi\alpha')^2}
 (g_{rr}+\parts x^ig_{ij}\parts x^j)\right)
+\frac{g^{\xi\xi}}{g^{t\xi}}
+
\frac{g^{rr}}{g^{t\xi}p_\xi} (p_j\parts
x^j+p_\xi \parts \xi)\parts \xi
\nonumber \\
\end{eqnarray}
and
\begin{eqnarray}
\partt x^i=\pb{x^i,H_{fix}}=\frac{1}{g^{t\xi}p_\xi}
(\parts x^ig^{rr}(p_j\parts x^j+ p_\xi
\parts \xi)+g^{ij}p_j) \ .
\nonumber \\
\end{eqnarray}
Further the equations of motion for
$p_\xi$ and $p_i$ take the form
\begin{eqnarray}
\partt p_\xi&=&\pb{p_\xi,H_{fix}}=
\parts \left[\frac{g^{rr}}{g^{\xi\xi}}
(p_j\parts x^j+p_\xi \parts \xi
)\right] \ ,
\nonumber \\
\partt p_i&=&
\pb{p_i,H_{fix}}= \parts
\left[\frac{g^{rr}} {g^{t\xi}p_\xi} p_i
(p_j
\parts x^j+p_\xi \parts \xi)+
\frac{g_{ij}\parts x^j}{g^{t\xi}
(2\pi\alpha')^2 p_\xi} \right] \ .
 \nonumber \\
\end{eqnarray}
In order to find the string configuration
that moves with constant velocity along
$\xi$ direction we consider 
following ansatz
 $ x^i=x^i(\sigma) \ , 
\partt \xi=v_\xi$. Then (\ref{eqxi})
take the form
\begin{eqnarray}
v_\xi
=-\frac{g_{tt}}{2 g_{t\xi}}
-\frac{g_{t\xi}} {2 (2\pi\alpha')^2
p_\xi^2} (g_{rr}+g_{ij}\parts x^i
\parts x^j)
\nonumber \\
\end{eqnarray}
that allows us to express $p_\xi$ as
\begin{equation}\label{pxi}
p_\xi=\pm\frac{1}{2\pi\alpha'}
\sqrt{-\frac{g_{t\xi}(g_{rr}+ g_{ij}
\parts x^i\parts x^j)}
{2v_\xi+\frac{g_{tt}}{g_{t\xi}}}} \ .
\nonumber \\
\end{equation}
On the other hand  the equation of
motion for $x^i$ is automatically
satisfied and the equation of motion
for $p_\xi$ implies $\partt p_\xi=0$.
Since we also consider the static
configuration we demand that $\partt
p_i=0$ so that the equation of motion
for $p_i$ implies
\begin{eqnarray}\label{pieq}
\frac{g_{ij}g_{t\xi} \parts x^j}{
p_\xi}=C' \nonumber \\
 \end{eqnarray}
using the fact that $p_i=0$.
 Then if we consider one mode $x\equiv x^1$
and insert (\ref{pxi}) into (\ref{pieq}) we
find the differential equation for $x$
\begin{eqnarray}
 \parts x=\pm\frac{C'}{2\pi\alpha'}
 \frac{\sqrt{g_{rr}}}
 {\sqrt{g_{xx}^2(-g_{tt}-2v_\xi
 g_{t\xi})-(\frac{C'}{2\pi\alpha'})^2
 g_{xx}}} \nonumber \\
\end{eqnarray}
We see that this equation exactly
reproduces the equation derived in
section (\ref{second}) after trivial
identification of the constants $C$ and
$C'$.

\end{appendix}

\end{document}